\documentclass[aps,prd,twocolumn,superscriptaddress,amssymb,eqsecnum,showpacs,showkeyes,secnumarabic,graphics,floatfix,nofootinbib,tightenlines,longbibliography]{revtex4-1}

\usepackage{graphicx}
\usepackage{epsfig}
\usepackage{bm}
\usepackage{cancel}
\usepackage{dcolumn}
\usepackage{amsmath}
\usepackage{hhline}
\usepackage[utf8]{inputenc}
\usepackage[breaklinks=true,colorlinks=true,
linkcolor=blue,urlcolor=blue,citecolor=cyan,
bookmarks=true,bookmarksopenlevel=2]{hyperref}

\def \beq  {\begin{equation}}
\def \eeq  {\end{equation}}
\def \ber  {\begin{eqnarray}}
\def \eer  {\end{eqnarray}}

\begin{document}
\newcommand{\newc}{\newcommand}

\newc{\be}{\begin{equation}}
\newc{\ee}{\end{equation}}
\newc{\ba}{\begin{eqnarray}}
\newc{\ea}{\end{eqnarray}}
\newc{\bea}{\begin{eqnarray*}}
\newc{\eea}{\end{eqnarray*}}
\newc{\D}{\partial}
\newc{\ie}{{\it i.e.} }
\newc{\eg}{{\it e.g.} }
\newc{\etc}{{\it etc.} }
\newc{\etal}{{\it et al.}}
\newc{\lcdm}{$\Lambda$CDM }
\newcommand{\nn}{\nonumber}
\newc{\ra}{\Rightarrow}

\title{Gravitational Interactions of Finite Thickness Global Topological Defects with Black Holes}
\author{L. Perivolaropoulos}\email{leandros@uoi.gr} 
\affiliation{Department of Physics, University of Ioannina, 45110 Ioannina, Greece}

\date {\today}

\begin{abstract}
It is well known that global topological defects induce a repulsive gravitational potential for test particles. 'What is the gravitational potential induced by black holes with a cosmological constant (Schwarzschild-de Sitter (S-dS) metric) on finite thickness global topological defects?'. This is the main question addressed in the present analysis. We also discuss the validity of Derrick's theorem when scalar field configurations are embedded in non-trivial gravitational backgrounds. In the context of the above stated question, we consider three global defect configurations: a finite thickness spherical domain wall with a central S-dS black hole, a global string loop with a S-dS black hole in the center and a global monopole near a S-dS black hole. Using an analytical  model, numerical simulations of the evolving spherical wall and energetic arguments we show that the spherical wall experiences a repelling gravitational potential due to the mass of the central black hole. This potential is further amplified by the presence of a cosmological constant. For initial domain wall radius larger than a critical value, the repulsive potential dominates over the wall tension and the wall expands towards the cosmological horizon of the S-dS metric where it develops ghost instabilities (the kinetic term changes sign). For smaller initial radius, tension dominates and the wall contracts towards the black hole horizon where it also develops ghost instabilities. We also show, using the same analytical  model and energetic arguments that a global monopole is gravitationally attracted by a black hole while a cosmological constant induces a repulsive gravitational potential as in the case of test particles. Finally we show that a global string loop with finite thickness experiences gravitational repulsion due to the cosmological constant which dominates over its tension for a radius larger than a critical radius leading to an expanding rather than contracting loop.
\end{abstract}
\maketitle

\section{Introduction}
\label{sec:Introduction}

It is well known that a non-trivial background metric has a significant effect of the dynamical equations determining the evolution of scalar fields. For example a non-trivial spherically symmetric background metric of the form
\be
ds^2= f(r) dt^2 - f(r)^{-1} dr^2 - r^2 (d\theta^2 +\sin^2\theta d\phi^2)
\label{sphmetric}
\ee
leads to a modified Klein-Gordon equation of the form
\be
\biggl[-\frac{r^{2}}{f(r)}\frac{\partial^{2}}{\partial t^{2}}+\frac{\partial}{\partial r}\biggl(r^{2}f(r)\frac{\partial}{\partial r}\biggr)-\mathbf{L}^{2}-m^{2}r^{2}\biggr]\Phi=0\ ,
\label{kgeq1}
\ee
where $\mathbf{L}$ is the angular momentum operator in spherical coordinates. Eq. (\ref{kgeq1}), its generalization for axisymmetric backgrounds  and the corresponding Dirac equation have been well studied \cite{Rowan:1976ug,Rowan:1977zg,Page:1976jj,Mukhopadhyay:1999gf,Elizalde:1987yn,Elizalde:1988it,Bezerra:2013iha,Vieira:2016ubt,Sakalli:2016fif,Vieira:2017ngr,Kraniotis:2018zmh,Bezerra:2017hrb,Kraniotis:2016maw},  exact solutions have been found using separation of variables and physical implications have been investigated (normal modes, Hawking radiation etc).

In the presence of a nonlinear scalar field potential $V(\Phi)$ the corresponding generalized scalar field dynamical equations have been studied at a smaller extend and mainly in the context of the existence of stable static solutions. In a flat 3+1 dimensional background such solutions are not allowed by Derrick's theorem\cite{Derrick:1964ww} which states that in 3+1 dimensions, any finite energy initially static scalar field configuration with canonical scalar kinetic terms and non-negative potential energy is unstable and energetically favoured to shrink and collapse. In a curved background this instability and lack of static solutions has been shown to persist in specific cases (eg charged rotating black hole\cite{Radmore:1978ux,Palmer1}) but no general statement has been made for arbitrary gravitational background. In the present analysis we demonstrate (among other results) that a proper choice of $f(r)$ in the background metric can lead to static and perhaps to metastable solutions thus evading the conclusion of Derrick's theorem. 

It has been shown that it is possible to evade Derrick's theorem\cite{Derrick:1964ww,Perivolaropoulos:1992kf} even in flat space and construct static topologically stable (or nontopological metastable\cite{Hindmarsh:1992yy,Achucarro:1992hs,James:1992wb,Vachaspati:1992fi,James:1992zp,Perivolaropoulos:1994zp,Achucarro:1999it}) scalar field solutions. Such approaches include the introduction of gauge fields\cite{Nielsen:1973cs,tHooft:1974kcl,Polyakov:1974ek,Perivolaropoulos:1993uj,Perivolaropoulos:2000hh,Perivolaropoulos:1993gg,Perivolaropoulos:2000hh}, the consideration of stationary rather than static solutions\cite{Coleman:1985ki,Kusenko:1997ad,Axenides:1999hs,Perivolaropoulos:1994zp,Copeland:1995fq} and the violation of the finite energy assumption made\cite{Bazeia:2007df,Bazeia:2003qt,Axenides:1999hs,Perivolaropoulos:1991du} with the possible introduction of non-standard kinetic terms\cite{Babichev:2006cy} in 3+1 or in higher dimensions\cite{Olasagasti:2000gx,Gregory:1999gv}.  A scalar field configuration with diverging energy in 3+1 dimensions would require a large scale cutoff which is naturally present in many physical systems. For example in a cosmological setup the role of the cutoff can be played by the horizon while in a condensed matter system the cutoff scale would be the size of the system. Global topological defects\cite{Vilenkin:1984ib,Vilenkin:2000jqa,Brandenberger:1993by,Hindmarsh:1994re,Sakellariadou:2006qs} in three spatial dimensions constitute such stable scalar field configurations with diverging energy and have observable effects in both condensed matter systems\cite{Mermin:1979zz,Chuang:1991zz,Digal:1998ak,Zurek:1996sj,Bowick:1992rz} and in cosmology\cite{Durrer:2001cg,Pen:1997ae,Vilenkin:1981zs,Brandenberger:1993by,Bazeia:2005jq,Perivolaropoulos:2005wa,Sikivie:1982qv,Lukas:1998yy,Harari:1987ht,Kusenko:1997si,Enqvist:1997si,Durrer:1998rw,Pen:1997ae,Bennett:1990xy,Perivolaropoulos:1992tz,Pogosian:1999np,Perivolaropoulos:1989ug,Perivolaropoulos:1992if,Moessner:1993za}. They include global monopoles \cite{Barriola:1989hx,Harari:1990cz,Hiscock:1990ev,Dando:1997gx,Barros:1997fi} (spherical field configurations) global strings \cite{Harari:1988wa,Vilenkin:1981kz,Cohen:1988sg} (axial field configurations) and domain walls \cite{Vilenkin:1981zs,Sikivie:1982qv,Csaki:2000wz,Gremm:2000dj,Gremm:1999pj} (planar configurations). 

Global defects constitute regions of physical space that may form during phase transitions in the Early Universe where vacuum energy of an early symmetric phase gets trapped for topological reasons. Unlike their gauged counterparts\cite{Hindmarsh:1994re}, global defects can not be approximated as being infinitely thin since the scalar field approaches its vacuum expectation value as a power law rather than exponentially. Thus they generically have a core of finite thickness of the order of the symmetry breaking scale that gave rise to the defects. This core may in general have non-trivial field structure with interesting effects in cosmology and condensed matter systems\cite{Axenides:1997ja,Axenides:1997sk}.

The evolution of topological defects and especially strings in curved backgrounds has been investigated\cite{DeVilliers:1998xz,DeVilliers:1997nk,Lonsdale:1988xd,Dabrowski:1996ph,Dubath:2006vs,Page:1998ya,deVega:1993hq,Frolov:1988zn,Natario:2017szw,Christensen:1998hg,Porfyriadis:1997re,Larsen:1994gh} under the thin defect approximation where the defects are assumed to have zero thickness and thus propagate via simplified forms of the action. 
For example the action that describes the evolution of a zero thickness string is the Nambu-Goto action which is proportional to the area of the world-sheet of the string.
This is a good approximation for gauged defects but it is not applicable for global defects where the finite thickness and the field structure can not usually be ignored. The evolution of global defects in non-trivial gravitational backgrounds requires the solution of the full dynamical field equations derived by variation of the scalar field action with the appropriate background metric. Such analyses and numerical simulations have been performed in the context of homogeneous expanding FRW background metrics for global defects\cite{Ye:1989dr,Press:1989yh,Lopez-Eiguren:2017dmc,Hagmann:1990mj} and gauged strings\cite{Ye:1990na,Bevis:2006mj,Bevis:2007qz,Daverio:2015nva,Lopez-Eiguren:2017ucu,Hindmarsh:2017qff,Urrestilla:2007yw} but not (to our knowledge) in inhomogeneous, spherically symmetric or axisymmetric black hole metrics.

Thus, an intersting question that needs to be adressed is the following: 'Is there an attractive interaction between black holes and global defects?'. How does the answer change if a cosmological constant is also present?'. Previous studies invetigating the global defect metric, have indicated that due to their vacuum energy, global topological defects induce a repulsive gravitational potential on test particles\cite{Harari:1988wa,Harari:1990cz,Vilenkin:1981zs,Ipser:1983db} in addition to a deficit angle. Based on this fact, a repulsive gravitational interaction between black holes and global defects could have been anticipated. On the other hand the equivalence principle\cite{DiCasola:2013iia} could imply that global defects would be attracted towards black holes. These apparently conflicting arguments motivate the more detailed study of the interaction of global defects with black holes and their evolution in black hole spacetimes.  Such a study is performed in this paper. We focus on particular global defect configurations with symmetric geometries. We use both an analytical  model based on energetic arguments and  numerical simulations of scalar field evolution which confirm the qualitative conclusions of the analytical model analysis.

In particular, we consider the following global defect configurations:
\begin{enumerate}
\item
A spherical domain wall with a Schwarzschild-de Sitter (S-dS) black hole at its center. We investigate analytically (using an approximate analytical model) and numerically (simulation) the evolution of the domain wall. Thus, we test both approaches by verifying the agreement of their results. We focus on identifying the gravitational interaction potential which is superposed with the spherical wall tension. 
\item
A global monopole at a given distance from a S-dS black hole. Using the above  analytical  model we investigate the gravitational potential that describes the evolution of the monopole in the S-dS background.
\item
A circular global string loop with a S-dS black hole at its center.  We use the same analytical  model tested in the case of the spherical wall to derive the gravitational potential that describes the evolution of the loop in the S-dS spacetime.
\end{enumerate}
In all the above cases we ignore the gravitational backreaction of the global defect on the S-dS spacetime.

The structure of this paper is the following: in section II we review the field dynamical equation describing the evolution of scalar fields in a nontrivial background metric as well as the energy of such field configuration. We also generalize the derivation of Derrick's theorem in a nontrivial gravitational background and demonstrate the possibility of evading this theorem in a gravitational background with specific properties. The case of scalar fields in a S-dS background metric is discussed in some detail. In section III we consider a spherical domain wall with degenerate vacua in and out of the sphere. A gravitational background metric corresponding to a S-dS black hole in the center of the sphere is assumed. The dynamical evolution of the spherical wall is analysed starting from an initially static configuration using an analytical  model and numerical simulations of the field evolution. In Section IV we implement the analytical model introduced and tested in section III to derive the gravitational interaction potential between a central S-dS black hole and a circular string loop or with a global monopole situated beyond the black hole horizon. Finally is section V we conclude summarize our basic results and diascuss implications as well as prospects for extensions of this project. Throughout this analysis we set $G=c=1$. We also rescale spacetime and mass scales to dimensionless form by using the scale of symmetry breaking $\eta$ of the global topological defects.

\section{Scalar Field Evolution in a non-trivial background metric and implications for Derrick's theorem}
\label{sec:Section 2}

The action describing the evolution of a canonical scalar field in a bacground metric $g_{\mu\nu}$ is of the form
\be
S=\int \left( g^{\mu\nu}\frac{\partial \Phi}{\partial x^\mu}\frac{\partial \Phi}{\partial x^\nu}-V(\Phi)\right) \sqrt{-g}\; d^4 x
\label{sflang}
\ee
where $g$ is the metric tensor determinant. Its variation leads to the dynamical field equation
\be
\frac{1}{\sqrt{-g}}\frac{\partial}{\partial x^\mu}g^{\mu \nu}\sqrt{-g}\frac{\partial \Phi}{\partial x^\nu}=-\frac{1}{2}V'(\Phi)
\label{fieldeq1}
\ee
where the ' denotes derivative. Using the isotropic metric (\ref{sphmetric}), the dynamical eq. (\ref{fieldeq1}) takes the form
\be
\frac{1}{f(r)}\frac{\partial^2}{\partial t^2}-\frac{1}{r^2}\frac{\partial}{\partial r}\left(r^2 f(r)\frac{\partial \Phi}{\partial r}\right)=-\frac{1}{2}V'(\Phi)
\label{fieldeq2}
\ee
where we have assumed a spherically symmetric field configuration. The energy of such a static field configuration is
\begin{widetext}
\be
E=\int d^3x \sqrt{-g}\; T_0^0=4\pi\int_{r_1}^{r_2} \left[r^2 f(r)\left(\frac{d\Phi}{dr}\right)^2+V(\Phi) r^2\right]dr
\label{energy1}
\ee
\end{widetext}
where $f(r)\geq 0$ and $r_1$ and $r_2$ correspond to the radial coordinates of the horizons ($f(r_i)=0$, $i=1,2$) of the metric (\ref{sphmetric}) and for a flat space they take 
the values $r_1=0$ and $r_2=+\infty$. For a S-dS metric
\be
f(r)=1-\frac{2m}{r}-\frac{\Lambda}{3}r^2
\label{S-dSfr}
\ee
and $r_1$ corresponds to the black hole horizon while $r_2$ corresponds to the cosmological horizon and $f(r)>0$ between the horizons.
The regions beyond the horizons ($r>r_2$ or $r<r_1$) are causally disconnected and correspond to ghost instabilities due to the change of sign of the kinetic term. Thus the integration is performed within the causally connected region between the horizons.

According to arguments based on Derrick's theorem, in flat space ($f(r)=1$), equation (\ref{fieldeq2}) does not have a finite energy static solution for $V(\Phi)\geq 0$. An intersting question to address is 'Is there a physically interesting form of $f(r)$ for which Derrick's theorem is evaded and thus there may be static solutions to the dynamical field equation (\ref{fieldeq2})?' 
\begin{figure}[!t]
\centering
\vspace{0cm}\rotatebox{0}{\vspace{0cm}\hspace{0cm}\resizebox{0.49\textwidth}{!}{\includegraphics{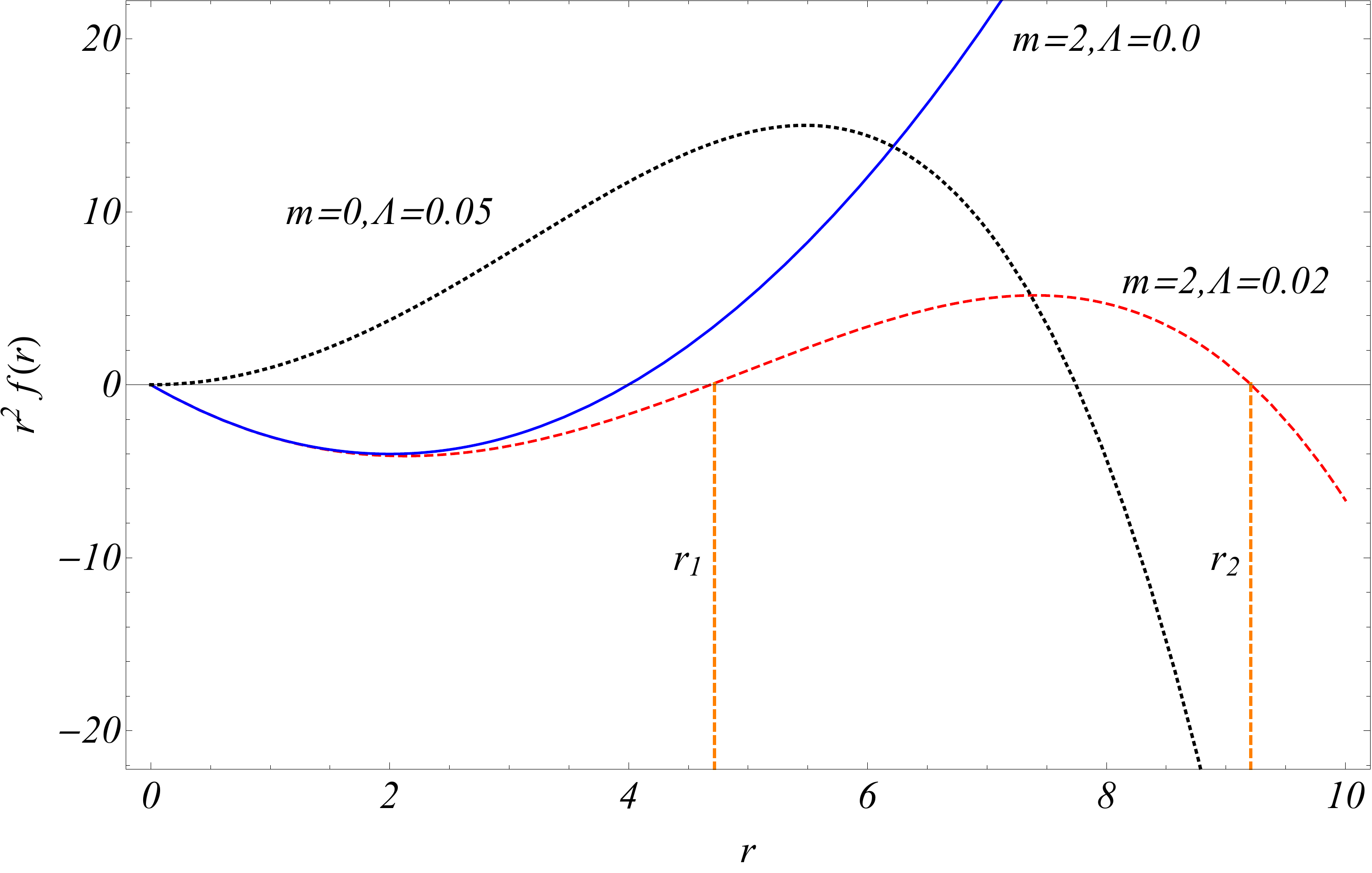}}}
\caption{The function $r^2 f(r)$ for a S-dS background metric is a decreasing function for a range of $r$ between the two horizons. Similarly for the de Sitter metric ($m=0$). In contrast for the Schwarzschild $f(r)$ is monotonically increasing.}
\label{figure1}
\end{figure}

In order to address this question we may consider an initially static field configuration $\Phi(r)$ and its rescaled form $\Phi_\alpha\equiv \Phi(\alpha r)$. We now search for an extremum and preferably a minimum of the energy functional (\ref{energy1}) with respect to the scaling parameter $\alpha$. Let $E_\alpha$ be the energy of the rescaled field configuration
\be
E_\alpha= 4\pi\int_{r_1}^{r_2} \left[r^2 f(r)\left(\frac{d\Phi_\alpha}{dr}\right)^2+V(\Phi_\alpha) r^2\right]dr
\label{energy2}
\ee
Using a change of variable $r'\equiv \alpha r$ and the facts $f(r_1)=f(r_2)=0$, $V(\Phi(r_1))=V(\Phi(r_2))=0$ it is straightforward to show that for the existence of a static solution a necessary condition is 
\be
\frac{1}{4\pi} \frac{dE}{d\alpha}\bigg\rvert_{\alpha=1}=I_1+I_2+I_3=0
\label{dedaeq1}
\ee
where
\ba 
I_1 &=& -\int_{r_1}^{r_2} r'^3 f'(r)\left(\frac{d\Phi}{dr}\right)^2  dr   \label{i1} \\
I_2 &=& -\int_{r_1}^{r_2} r^2 f(r)\left(\frac{d\Phi}{dr}\right)^2  dr  \label{i2} \\
I_3 &=& -\int_{r_1}^{r_2} r^2 \;V(\Phi)\;  dr  \label{i3} 
\ea
Since $I_3<0$ and $I_2<0$ we need $I_1>0$ in order to satisfy eq. (\ref{dedaeq1}) and have a static solution. Thus, we need $f'(r)<0$ at least for some range between the horizons. This condition can not be satisfied in a flat space ($f(r)=1$) and this is consistent with Derrick's theorem. It is also not satisfied in a Schwarzschild metric ($f(r)=1-\frac{2m}{r}$) where $f(r)$ is a monotonically increasing function (Fig. \ref{figure1}). Thus Derrick's theorem is also applicable for this metric (no static solution exists). A similar argument\cite{Radmore:1978ux,Palmer1} exists for charged Reissner–Nordström black holes where 
\be
f(r)=f(r)=1-\frac{2m}{r}+\frac{e^2}{r^2}
\label{rnbhfr}
\ee
In this case $r_1=m+\sqrt{m^2+e^2}$ and $r_2=+\infty$ and as in the Schwarzschild metric $f(r)$ is monotonically increasing in the integration range, leading to $I_1<0$ and no static solution exists. 

As shown in Fig. \ref{figure1}, a metric for which $f(r)$ is not monotonic is the S-dS metric (eq. (\ref{S-dSfr})). For this metric $f(r)$ has a maximum $f(r_{max})$ in the range $[r_1,r_2]$ and is a decreasing function for $r\in [r_{max},r_2]$. In this range $f'(r)<0$ and thus it is possible to have $I_1>0$ (static solution). Thus, Derrick's theorem is not applicable for this metric and a static solution is allowed to exist.  However such a solution is unstable for the spherical domain wall in S-dS background configuration as discussed in the next section. 

In general the stability of the static solution depends on the sign of the second derivative of the energy. The necessary condition for stability is 
\be
\frac{d^2E}{d\alpha^2}\bigg\rvert_{\alpha=1}>0
\label{ddea1}
\ee
The left hand side of this inequality depends on both the $f'(r)$ and $f''(r)$ and thus we anticipate that for proper choice of $f(r)$ it may be satisfied leading to metastable static scalar field configuration. We thus conclude that Derrick's theorem can be evaded in properly chosen nontrivial gravitational backgrounds.

\section{Spherical Domain Wall in a Schwarzschild-DeSitter Background}
\label{sec:Section 3}

We now focus on the particular class of scalar field potentials that correspond to spontaneous symmetry breaking and give rise to global topological defects. The simplest topological defect, the domain wall, may form in theories where the potential $V(\Phi)$ has a discrete set of degenerate minima. Such is the double well potential leading to a spontaneous breaking of $Z_2$ symmetry. It is of the form
\be 
V(\Phi)=\frac{1}{2} \left(\Phi^2-\eta^2\right)^2
\label{potdomwall}
\ee
where $\eta$ is the scale of symmetry breaking. A spherical domain wall is a field configuration that interpolates between the two degenerate minima $\pm\eta$ of the potential (\ref{potdomwall}) as the surface of the wall sphere in physical space is crossed. An example of such a spherically symmetric field configuration is
\be 
\Phi(r)=\eta \; \tanh\left[\eta\left(r-r_0\right)\right]
\label{phidomwall}
\ee
\begin{figure}[!t]
\centering
\vspace{0cm}\rotatebox{0}{\vspace{0cm}\hspace{0cm}\resizebox{0.4\textwidth}{!}{\includegraphics{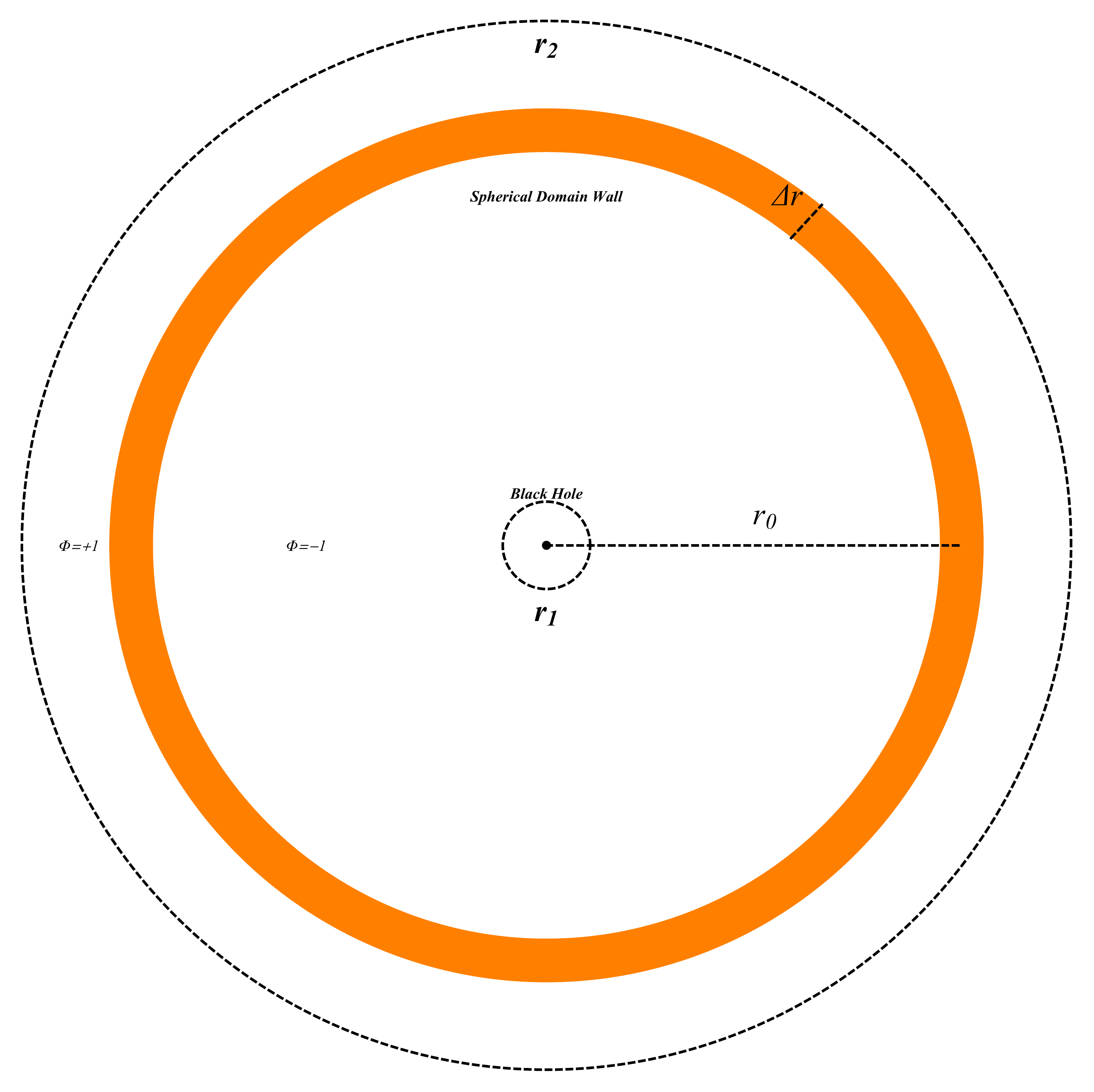}}}
\caption{The geometry of the spherical domain wall described by the ansatz (\ref{phidomwall}) with the presence of a S-dS black hole in its center. The horizons are also shown.}
\label{figure2}
\end{figure}

\begin{figure}[!t]
\centering
\vspace{0cm}\rotatebox{0}{\vspace{0cm}\hspace{0cm}\resizebox{0.49\textwidth}{!}{\includegraphics{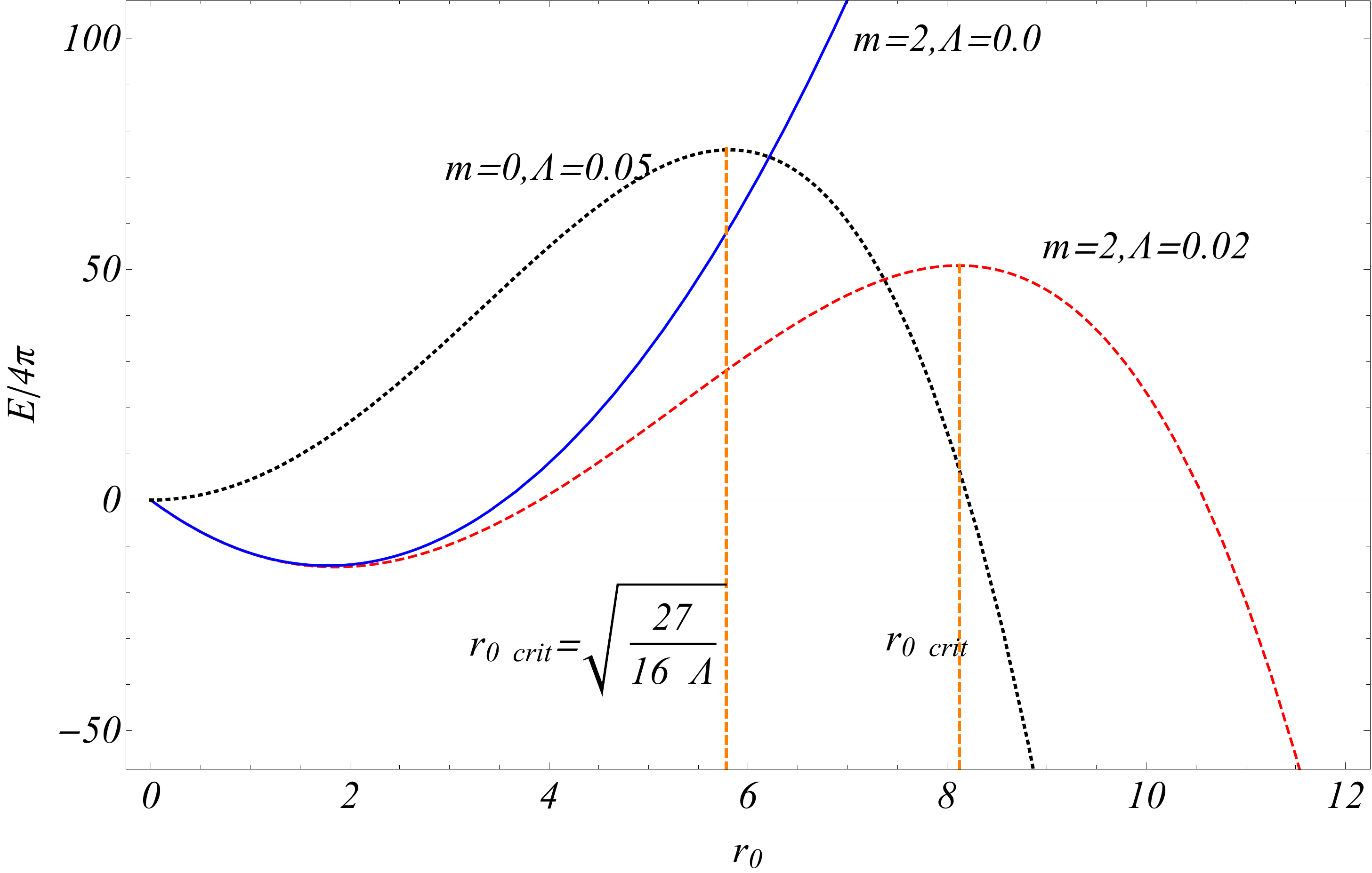}}}
\caption{An approximate form of the spherical domain wall energy as a function of its radius $r_0$ as obtained from eq. (\ref{energywallapprox1}).}
\label{fig3}
\end{figure}
The basic features of the evolution of this configuration describing a spherical domain wall may be obtained analytically via a simple analytical model based on an approximate form of the energy functional. They may also be obtained numerically by either explicit minimization of the full energy functional or by numerical simulation of the wall evolution by solving the dynamical equation (\ref{fieldeq2}) with the initial condition (\ref{phidomwall}).

\subsection{Analytical  Model}
\label{sec:Section 3a}

The only scale of the model in a flat background space is the scale of symmetry breaking $\eta$ which also describes both the width $\Delta r$ of the domain wall as
\be
\Delta r \simeq \eta^{-1}
\label{deltar}
\ee
and the scale of variation of the scalar field 
\be
\Delta \Phi = 2\eta
\label{deltaphi}
\ee
from one vacuum to the other. The geometry of the spherical domain wall described by the ansatz (\ref{phidomwall}) is shown in Fig. \ref{figure1}. 

In what follows we rescale spacetime coordinates and energy/mass scales by the scale of symmetry breaking $\eta$. We thus set $\eta=1$ and use the rescaled dimensionless quantities. 
\begin{figure*}[ht]
\centering
\begin{center}
$\begin{array}{@{\hspace{-0.10in}}c@{\hspace{0.0in}}c}
\multicolumn{1}{l}{\mbox{}} &
\multicolumn{1}{l}{\mbox{}} \\ [-0.2in]
\epsfxsize=3.3in
\epsffile{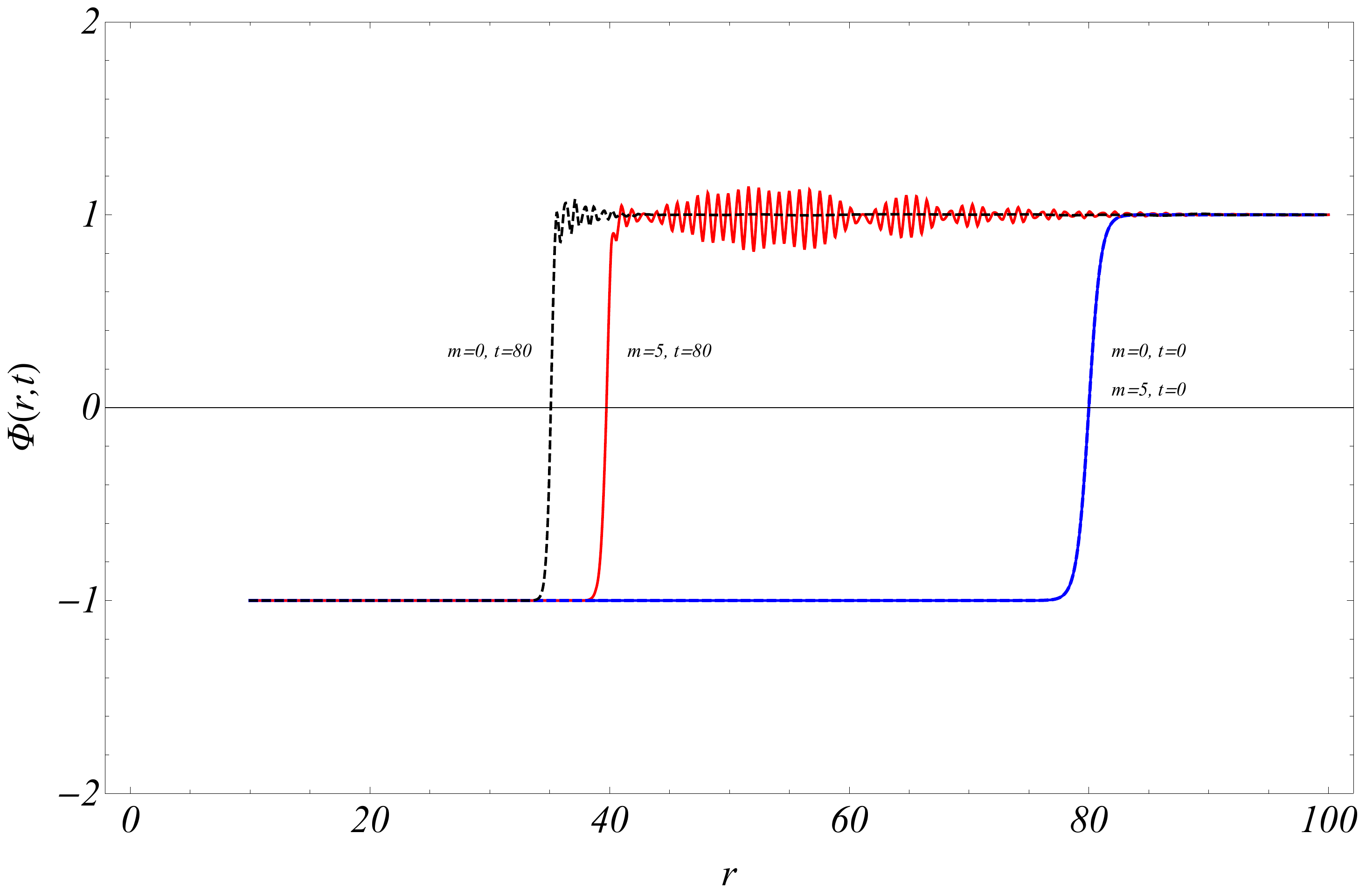} &
\epsfxsize=3.3in
\epsffile{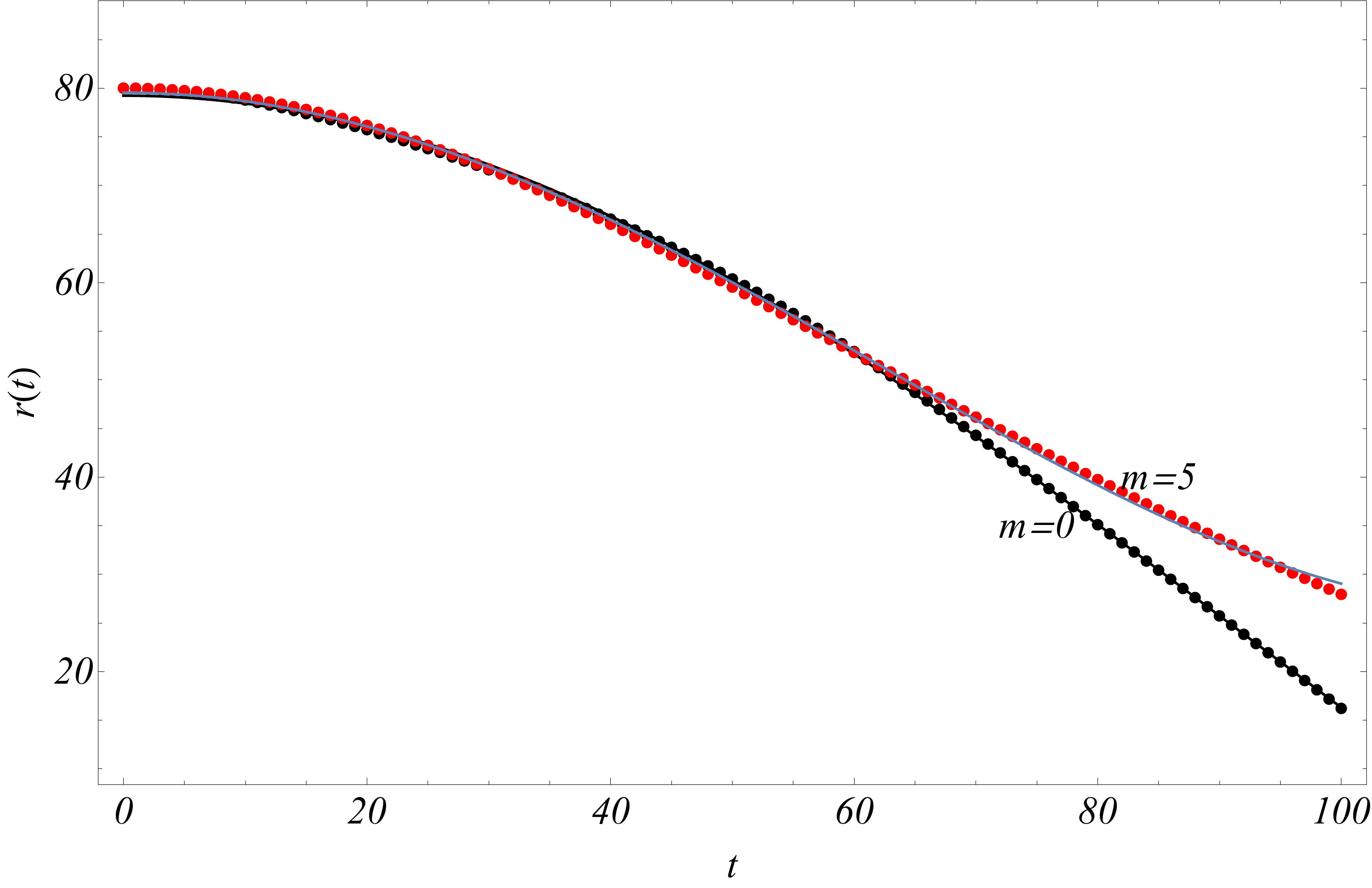} \\
\end{array}$
\end{center}
\vspace{0.0cm}
\caption{Left panel: Evolved spherical wall field configuration in flat space (black dotted line) and in the presence of a central black hole with $m=5$ $\Lambda=0$ (red line) at $t=80$. The blue line corresponds to the initial condition. Clearly the black hole delays the collapse of the wall as expected from the analytical approximate model. Right panel: The location of the zero of the scalar field (domain wall) as a function of time in the absence (black dots) and in the presence (red dots) of a central black hole. The continuous line in the black dots is a fit with a cosine harmonic function and the line in the red dots is a cosine+quadratic term in time. Due to the constant repulsive force, the black hole introduces a repulsive term quadratic in time that becomes apparent at late times.}
\label{fig4}
\end{figure*}

\begin{figure*}[ht]
\centering
\begin{center}
$\begin{array}{@{\hspace{-0.10in}}c@{\hspace{0.0in}}c}
\multicolumn{1}{l}{\mbox{}} &
\multicolumn{1}{l}{\mbox{}} \\ [-0.2in]
\epsfxsize=3.3in
\epsffile{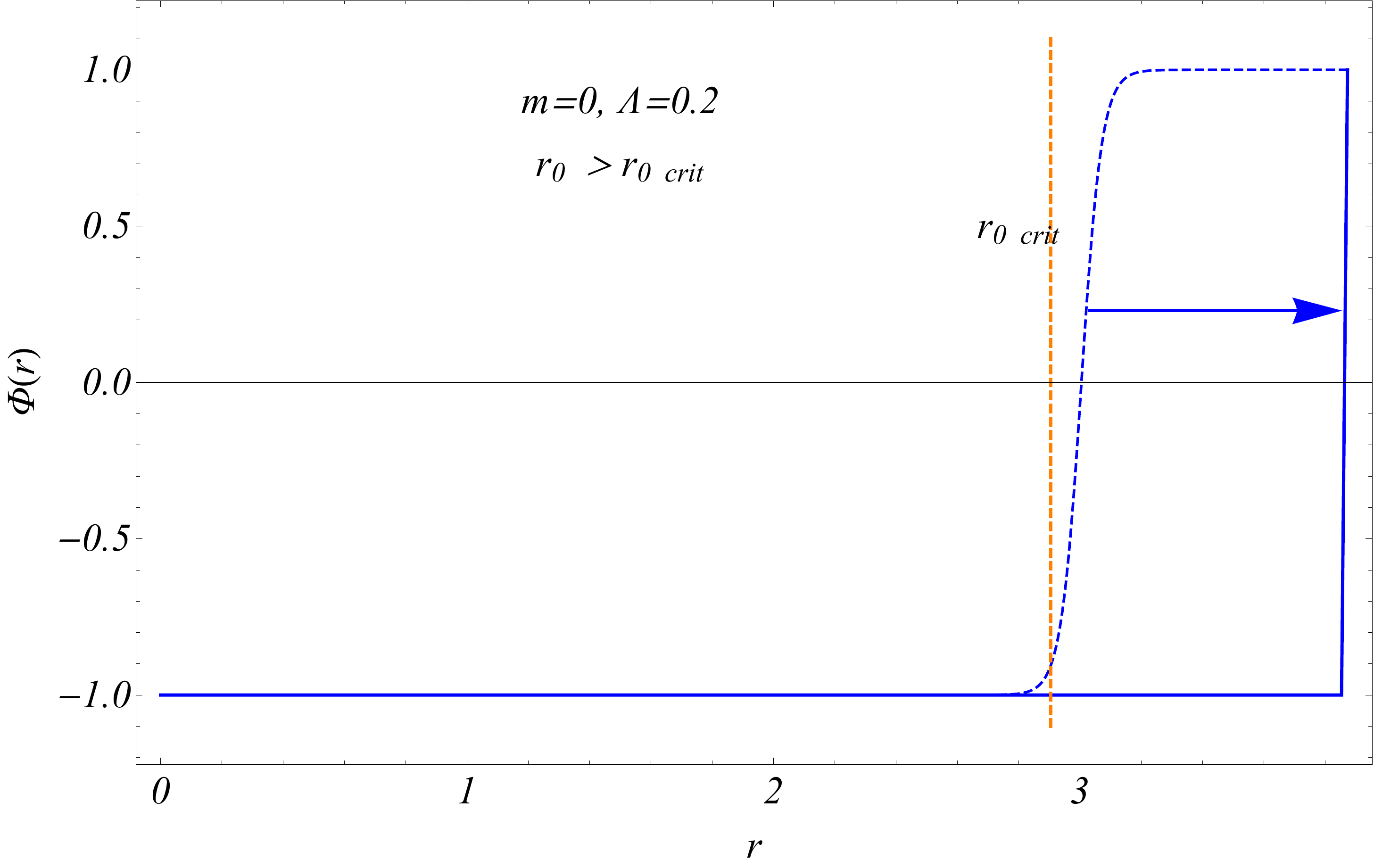} &
\epsfxsize=3.3in
\epsffile{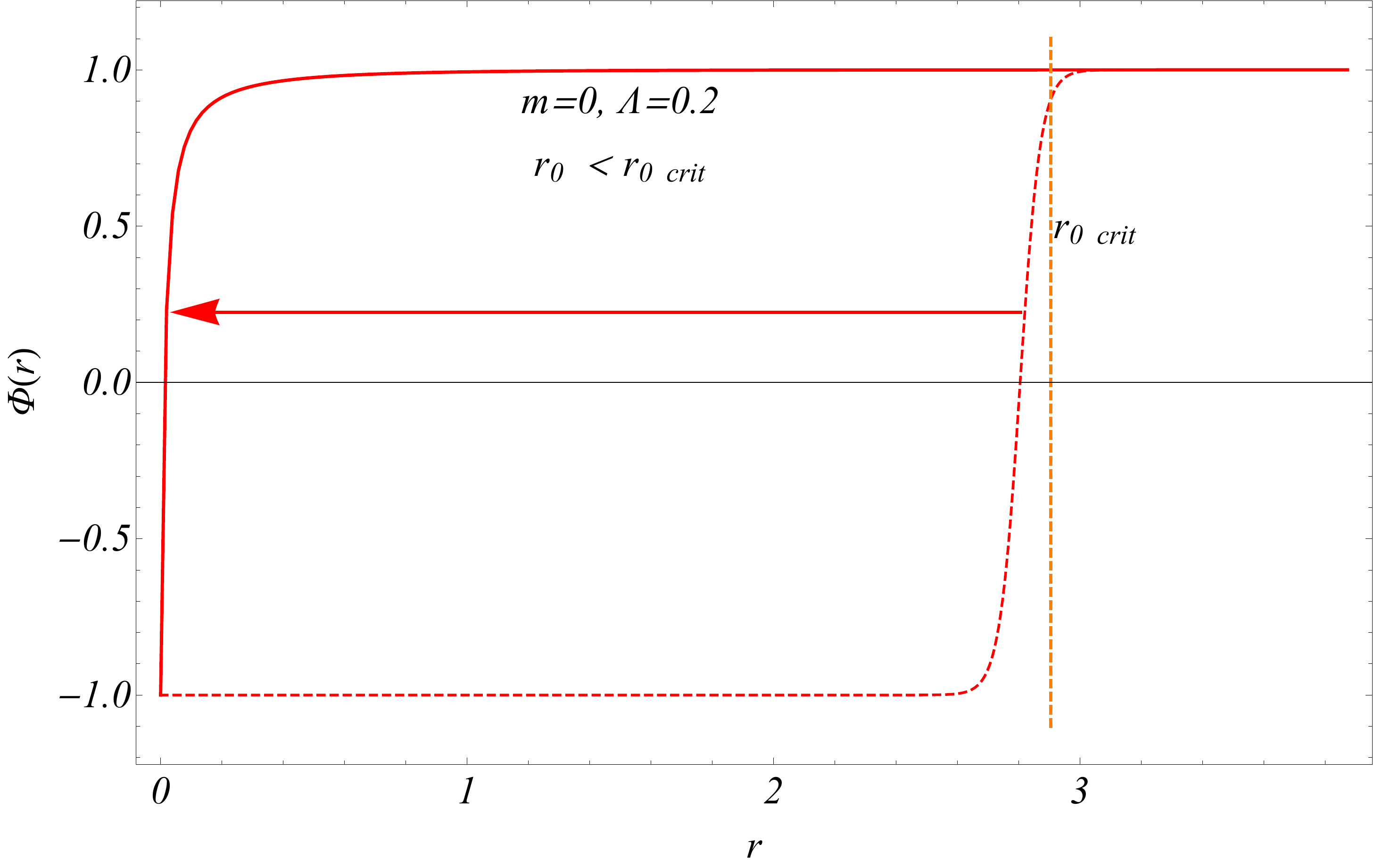} \\
\end{array}$
\end{center}
\vspace{0.0cm}
\caption{The field configuration corresponding to minimum energy is expanded to the outer boundary-horizon (left panel) if the initial guess wall radious $r_0$ is larger than the critical radius $r_{0crit}$ corresponding to the energy maximum of Fig. \ref{fig3}. Similarly, the minimum energy field configuration is collapsed to the inner boundary-horizon (right panel) if the initial guess wall radius $r_0$ is smaller than the critical radius $r_{0crit}$ corresponding to the energy maximum of Fig. \ref{fig3}. The energy minimization was performed numerically using 200 lattice points.}
\label{fig5}
\end{figure*}
The evolution of the spherical domain wall (\ref{phidomwall}) is described by the action (\ref{sflang}) and the corresponding dynamical equation (\ref{fieldeq2}).
The energy of the spherical wall, assumed initially static is given by eq. (\ref{energy1}). Assuming a small but finite thickness of the domain wall (Fig. \ref{figure2}) and a S-dS background metric we may approximate its energy as
\be 
\frac{E}{4\pi}\simeq \left(r_0^2-2mr_0-\frac{\Lambda}{3} r_0^4\right)\left(\frac{\Delta \Phi}{\Delta r}\right)^2 \Delta r +  V(0)\; r_0^2 \; \Delta r
\label{energywallapprox}
\ee
\begin{figure*}[ht]
\centering
\begin{center}
$\begin{array}{@{\hspace{-0.10in}}c@{\hspace{0.0in}}c}
\multicolumn{1}{l}{\mbox{}} &
\multicolumn{1}{l}{\mbox{}} \\ [-0.2in]
\epsfxsize=3.3in
\epsffile{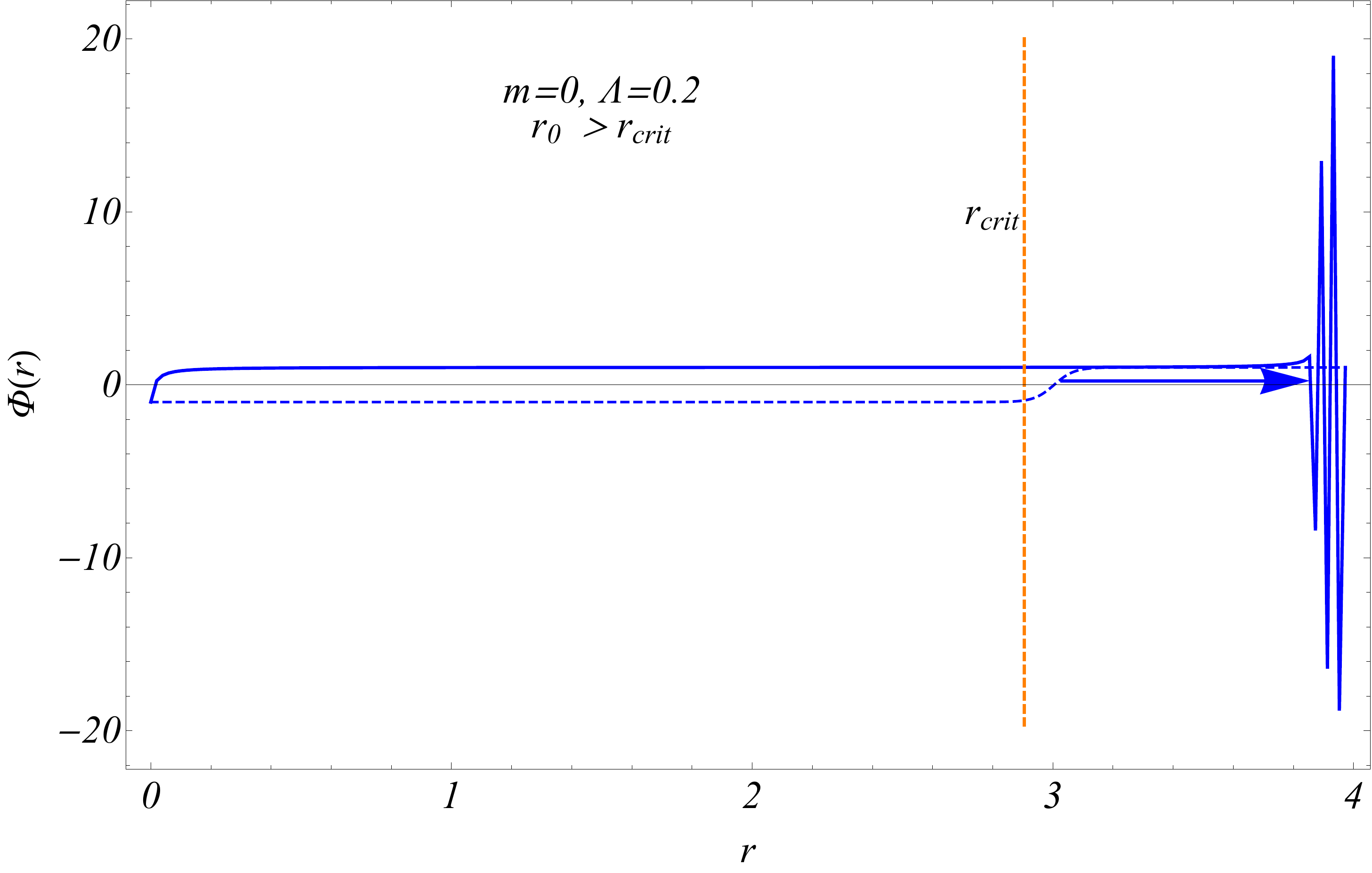} &
\epsfxsize=3.3in
\epsffile{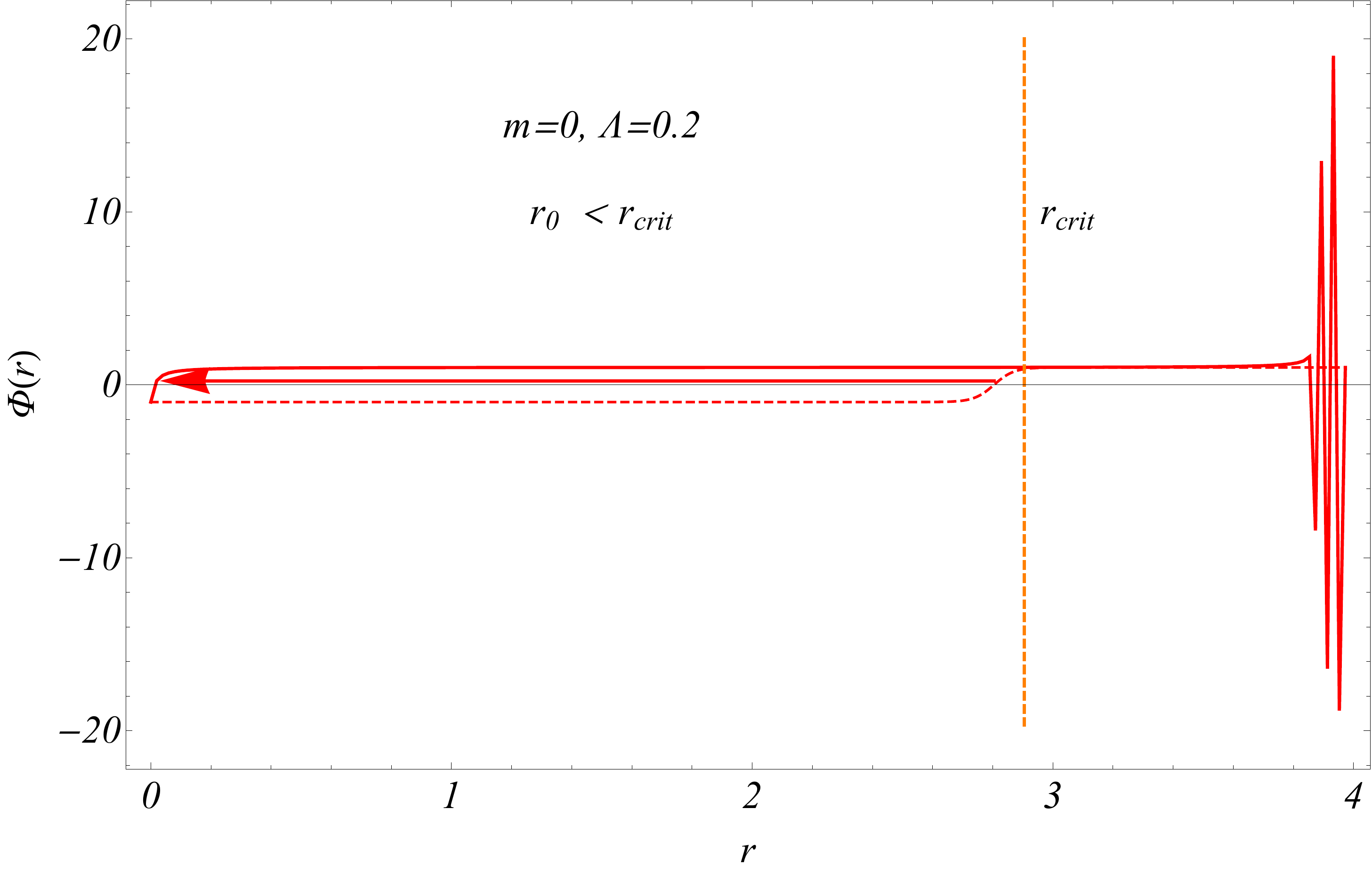} \\
\end{array}$
\end{center}
\vspace{0.0cm}
\caption{The field configuration emerging after energy minimizatio with boundary conditions imposed beyond the horizon. If the boundary conditions are imposed beyond the cosmological horizon the oscillating instabilities develop due to the negative kinetic term in the region beyond the horizon. These instabilities however do not affect the collasing/expanding properties of the field in the region where the gradient term is positive. Similar results are obtained if we use $m\neq 0$ and impose the inner boundary condition inside the black hole horizon.}
\label{fig6}
\end{figure*}
Using eqs (\ref{deltar}), (\ref{deltaphi}) in (\ref{energywallapprox}) with $\eta\rightarrow 1$ we find
\be 
\frac{E}{4\pi}\simeq \frac{9}{2}r_0^2-8\;m\;r_0-\frac{4}{3}\Lambda \; r_0^4
\label{energywallapprox1}
\ee
This approximate expression for the domain wall energy as a function of its radius $r_0$ is shown in Fig. \ref{fig3}

For small values of $r_0$ the repulsive term proportional to the black hole mass $m$ dominates leading to a decreasing energy. For intermediate values of $r_0$ the tension $\sim r_0^2$ dominates leading to an increasing energy while for large wall radius the repulsive term due to the cosmological $\Lambda$ dominates and the energy becomes again a decreasing function of $r_0$. For $m=0$ the energy is maximised for 
\be
r_{0crit}=\sqrt{\frac{27}{16\Lambda}}
\label{r0critm0}
\ee
For $r_0>r_{0crit}$ the wall is expected to expand due to the repulsive potentials of the cosmological constant and the black hole mass while for $r_0<r_{0crit}$ contraction is expected due to the tension. The existence of an unstable static solution is also anticipated for $r_0=r_{0crit}$. Clearly, the approximate expression for the energy (\ref{energywallapprox1}) can only by trusted in regions between the horizons where the coefficient of the gradient term remains positive.

The attractive/repulsive nature of each term is also seen via the total force applied on the domain wall due to the tension, the black hole mass and cosmological constant. This force is obtained as
\be
F\equiv -\frac{\partial E}{\partial r_0}=-9\; r_0+8\; m + \frac{16}{3}\; \Lambda\; r_0^3
\label{force1}
\ee
The above described predicted evolution of the spherical wall obtained using this simple analytical model is verified by numerical simulation of its evolution in the next subsection.

\subsection{Numerical Derivation of Domain Wall Evolution}
\label{sec:Section 3b}

The repulsive potential due to the mass of the central black hole predicted by the analytical  model may be verified numerically by simulating the domain wall evolution. Thus we solve the dynamical equation (\ref{fieldeq2}) for a S-dS background metric (\ref{S-dSfr}) and the symmetry breaking potential (\ref{potdomwall}). In order to focus on the effects of the mass term we first set $\Lambda=0$ in eq. (\ref{S-dSfr}).We use the initial field configuration of eq. (\ref{phidomwall})  corresponding to a spherical domain wall and boundary conditions at the two horizons
\ba  
\Phi(r_1)&=&-1 \label{bc1}\\
\Phi(r_2)&=&1 \label{bc2}
\ea
We evolve the configuration in time in both a flat background space ($m=0$) and in the presence of the black hole. The results are shown in  Fig. \ref{fig4}.
Clearly, the repulsive gravitational interaction originating from the black hole mass delays the collapse of the spherical wall due to its tension. The equivalence principle is not violated by this repulsive interaction since the spherical wall is a non-local object and thus the gravitational effects on it can not be eliminated at any frame.

A simple way to derive numerically the basic features of the evolution of the  domain wall initial configuration (\ref{phidomwall}) is to explicitly minimise the energy functional (\ref{energy1}) with fixed boundary conditions at the two horizons $r_1$, $r_2$ corresponding to the two distinct vacua of the potential (\ref{potdomwall}) (eqs (\ref{bc1})-(\ref{bc2})). We approximate the integral (\ref{energy1}) as a discrete sum over 200 lattice points and minimize with respect to the 200 values of the field (one value at each lattice point) starting from the initial configuration (\ref{phidomwall}). Thus the energy integral (\ref{energy1})  is written as
\be
E=dx\sum_{i=1}^N \left[r_i^2 \; f(r_i)\left(\frac{\Phi_i-\Phi_{i-1}}{dx}\right)^2 + r_i^2 V(\Phi_i)\right]
\label{edescrete}
\ee
where we have set $N=200$ and $V(\Phi_i)$ is given by eq. (\ref{potdomwall}). For simplicity we set $m=0$, $\Lambda=0.2$ in evaluating $f(r_i)$ from eq. (\ref{S-dSfr}) which implies that $r_1=0$ and $r_{0crit}\simeq 2.9$ as obtained from eq. (\ref{r0critm0})

As expected based on the above described analytical model and the energy shown in Fig. \ref{fig3}, the field configuration that minimizes the energy subject to the boundary conditions (\ref{bc1})-(\ref{bc2}) depends on the initial location of the domain wall (value of $r_0$ of the initial guess used for the minimization). For $r_0<r_{0crit}$ ($r_0>r_{0crit}$) the final field configuration corresponds to a domain wall collapsed (expanded) on the inner (outer) horizon at $r_1$ ($r_2$) where the boundary condition stops its further collapse (expansion). This effect is demonstrated in Fig. \ref{fig5} which shows the initial guess wall configuration and the final configuration emerging after the energy minimization. 

These results imply that there is an unstable static spherical domain wall solution for a radius $r_0=r_{0crit}$. The instability of this solution may also be seen by evaluating the second derivative of the energy (\ref{energywallapprox1}) with respect to $r_0$ and showing that as expected it is negative at the maximum $r_{0crit}$.

If the boundary conditions (\ref{bc1})-(\ref{bc2}) of the energy minimization are imposed beyond the two horizons where the field gradient term in the energy changes sign, then the expected oscillating instabilities become manifest in the regions beyond the horizons. These causally protected ghost instabilities are demonstrated in Fig. \ref{fig6}.

\begin{figure*}[ht]
\centering
\begin{center}
$\begin{array}{@{\hspace{-0.10in}}c@{\hspace{0.0in}}c}
\multicolumn{1}{l}{\mbox{}} &
\multicolumn{1}{l}{\mbox{}} \\ [-0.2in]
\epsfxsize=3.3in
\epsffile{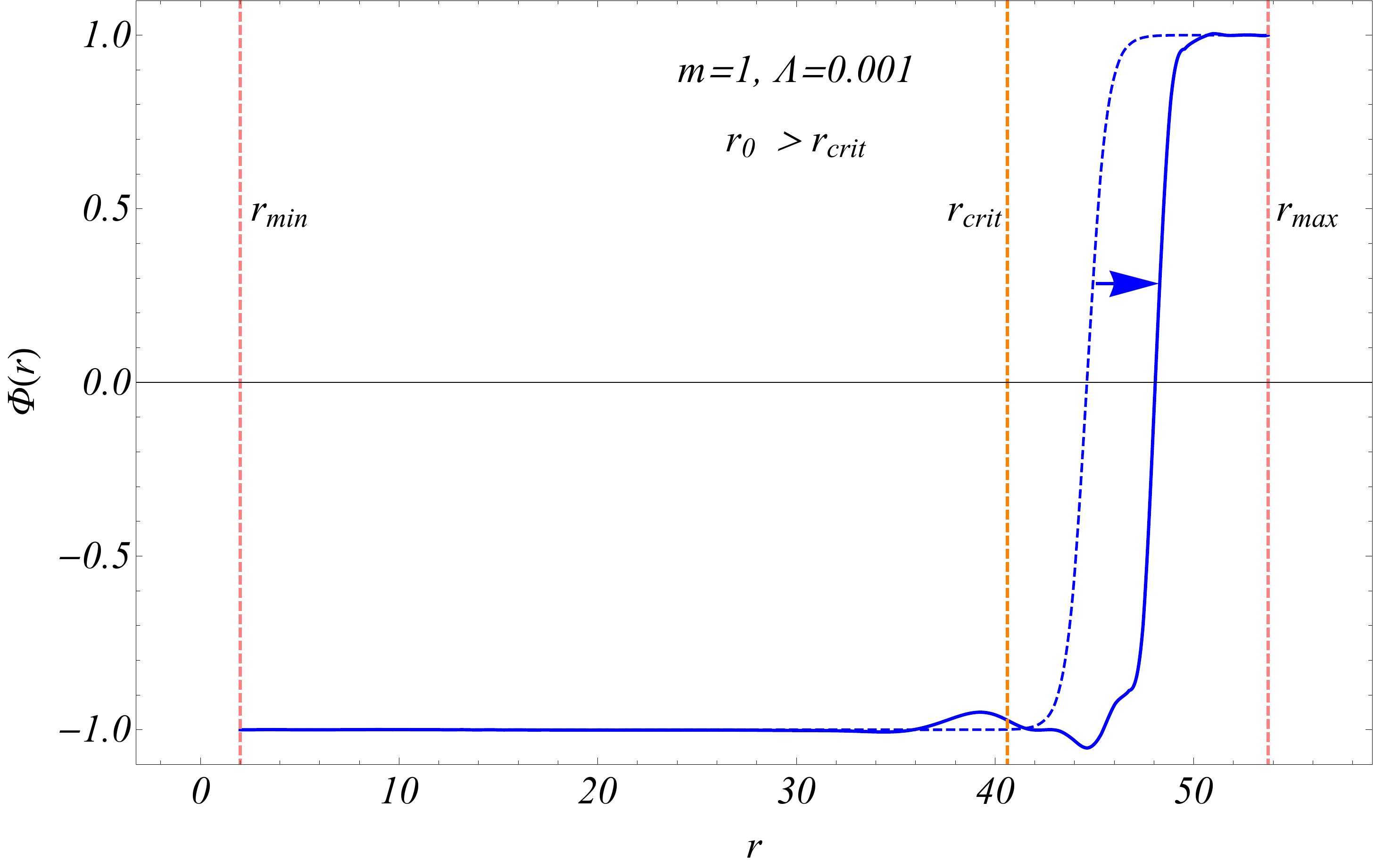} &
\epsfxsize=3.3in
\epsffile{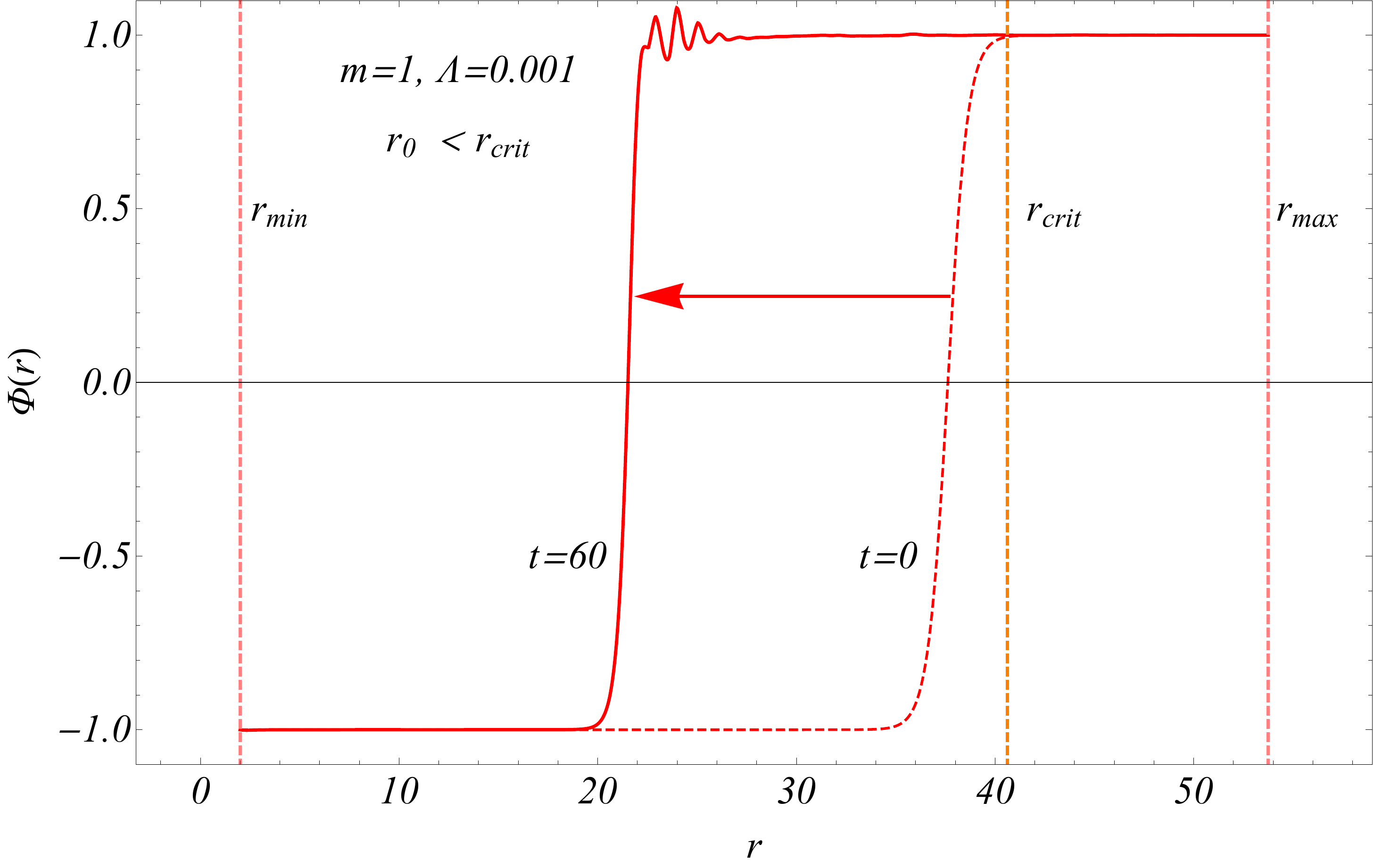} \\
\end{array}$
\end{center}
\vspace{0.0cm}
\caption{The numerical evolution of the spherical wall depends crucially on the initial conditions. If the initial radius $r_0$ is larger than the energy maximum of Fig. (\ref{fig3}) the wall expands while if the initial radius $r_0$ is smaller than the energy maximum of Fig. (\ref{fig3}) the wall contracts.}
\label{fig7}
\end{figure*}

The collapsing/expanding behavior of the spherical domain wall which depends on its initial radius may also be demonstrated by explicit numerical solution of the dynamical field equation (\ref{fieldeq2}) in the region between the horizons with boundary conditions (\ref{bc1})-{\ref{bc2}) and initial condition given by eq. (\ref{phidomwall}). As shown in Fig. \ref{fig7} the evolution of the spherical wall depends crucially on its initial radius $r_0$. If $r_0>r_{0crit}$ the wall expands to the outer horizon while if $r_0<r_{0crit}$ the wall contracts towards the inner horizon. Despite of the approximations used in the analytical derivation of $r_{0crit}$ from eq. (\ref{energywallapprox1})  we have found numerically that its value is accurate to within better than $5\%$. This is surprising in view of the significant approximations involved in deriving eq. (\ref{energywallapprox1}).

\section{Global monopole and global string loop in a S-dS spacetime}
\label{sec:Section 4}

\subsection{Global Monopole in a S-dS spacetime}
\label{sec:Section 4a}

The analytical  model introduced in section \ref{sec:Section 3a} has been demonstrated to describe the qualitative features of the evolution of the spherical domain wall in the presence of a non-trivial background metric in a satisfactory manner. Motivated by this result we apply the same model in this section to obtain the gravitational interaction potential between a global monopole and a S-dS black hole.
\begin{figure}[!t]
\centering
\vspace{0cm}\rotatebox{0}{\vspace{0cm}\hspace{0cm}\resizebox{0.49\textwidth}{!}{\includegraphics{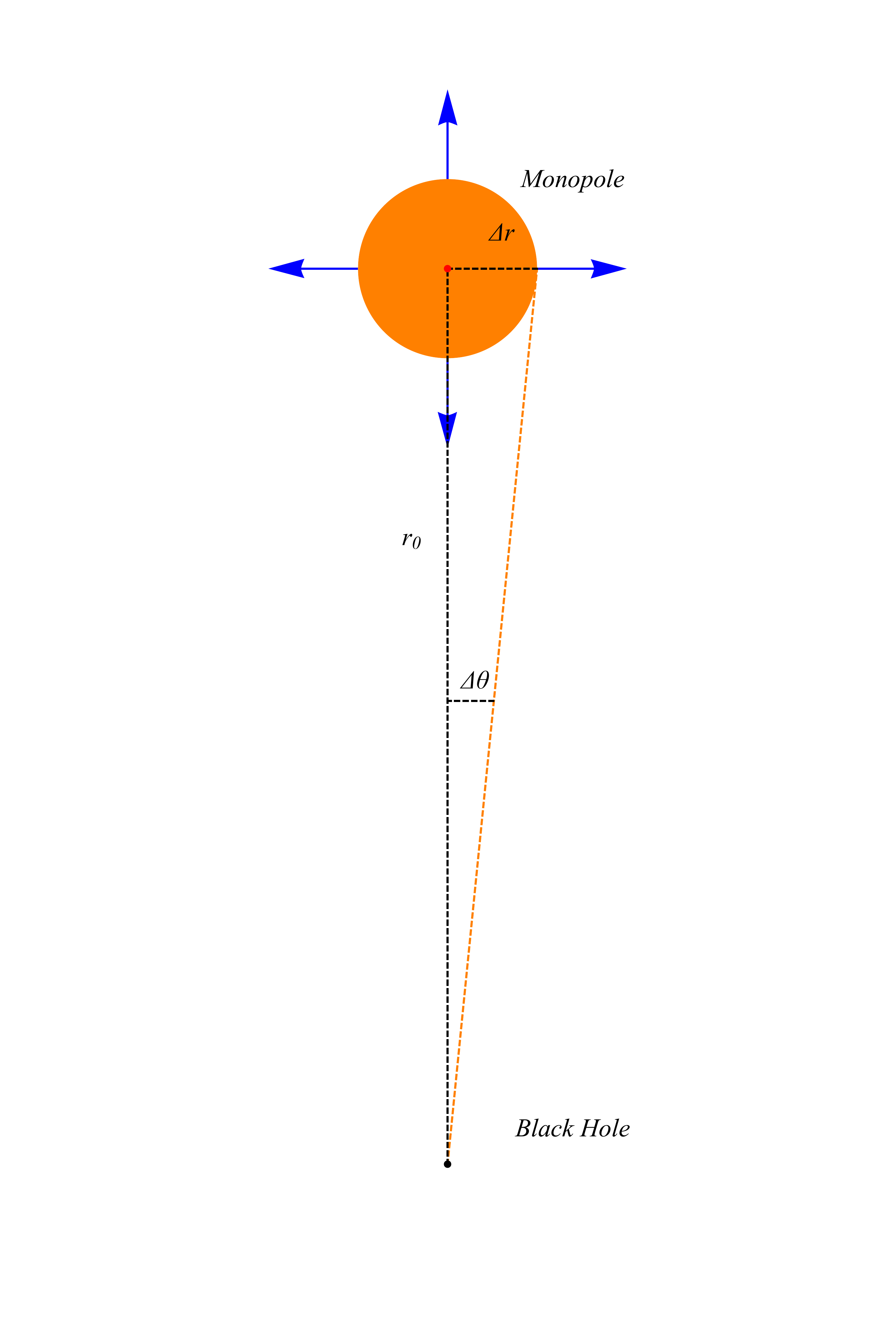}}}
\caption{The global monopole as viewed from the spherical  coordinate system with center on the black hole.}
\label{fig8}
\end{figure}
A global monopole with unit topological charge corresponds to a topologically stable, spherically symmetric hedgehog triplet scalar field configuration of the form
\be 
\Phi^a=\Phi(r) n^a
\label{monopconf}
\ee
where $a=1,2,3$, $n^a\equiv (\sin \theta \cos \varphi, \sin\theta \sin \varphi, \cos \varphi)$ and the boundary conditions for $\Phi(r)$ are
\ba
\Phi(r=0)&=&0 \label{bc1mon} \\
\Phi(r\rightarrow \infty)&=& 1 \label{bc2mon}
\ea
The action describing the dynamics of a global monopole is of the form
\be
S=\int \left( g^{\mu\nu}\frac{\partial \Phi^a}{\partial x^\mu}\frac{\partial \Phi^a}{\partial x^\nu}-V(\sqrt{\Phi^a \Phi^a})\right) \sqrt{-g}\; d^4 x
\label{sflangmon}
\ee
where $V(\sqrt{\Phi^a \Phi^a})=V(\Phi)$ is given by eq. (\ref{potdomwall}) and in this case leads to the breaking of a global $O(3)$ symmetry to $O(2)$.
The energy of the global monopole in a spherical coordinate system whose center coincides with the monopole center is
\begin{widetext}
\be
E=\int T_0^0\; \sqrt{-g} \; dr \;d\theta \; d\phi=
\int \left[-g^{rr}(\partial_r \Phi)^2-g^{\theta \theta}(\partial_\theta \Phi)^2-g^{\theta \theta}\Phi^2-g^{\varphi\varphi}(\partial_\varphi \Phi)^2 -g^{\varphi\varphi}\Phi^2 \sin^2\theta + V(\Phi)\right]\; \sqrt{-g} \; dr \;d\theta \; d\phi
\label{energymon}
\ee
\end{widetext}
The terms proportional to $\Phi^2$ originate from the angular gradients and lead to a contribution to the energy that is independent of $r_0$ and would be diverging in flat space. In the S-dS spacetime where the horizons provide natural cutofs the integral of these terms is proportional to a natural cutoff scale $R_{max}$ of the S-dS spacetime provided by cosmological S-dS horizon ie 
\begin{widetext}
\be 
E_{ang}=-\int (g^{\theta \theta}\Phi^2+g^{\varphi\varphi}\Phi^2 \sin^2\theta)\sqrt{-g} \; dr \;d\theta \; d\phi \simeq 2 \int \sin\theta \; dr \; d\theta \; d\phi \sim 8\pi R_{max}
\label{energymonang}
\ee
\end{widetext}
where $R_{max}$ is the cutoff scale of the order of the cosmological S-dS horizon $r_2$. The rest of the terms in the bracket of eq. (\ref{energymon}) may be written in the form $g^{ij}\partial_i\Phi \partial_j \Phi$ which is invariant under coordinate transformations. The total energy is also invariant under a change of the origin of the coordinate system used and thus we can shift the coordinate system from the monopole center to the black hole center keeping the same expression for the energy. This configuration is shown in Fig. \ref{fig8}. In this shifted coordinate system, the monopole field depends on both $r$ and $\theta$ but retains its azimouthal invariance (independence from $\varphi$). Thus the energy (\ref{energymon}) may be expressed as 
\begin{widetext}
\be 
E=2\pi \int_{r_1}^{r_2}\int_0^\pi \left[f(r)\left(\partial_r \Phi(r,\theta)\right)^2+\frac{1}{r^2}\left(\partial_\theta \Phi (r,\theta)\right)^2+\frac{2}{r^2}\Phi (r,\theta)^2+V(\Phi)\right]r^2 \sin \theta dr\; d\theta
\label{energymon1}
\ee
\end{widetext}
where $f(r)$ corresponds to the S-dS metric (eq. (\ref{S-dSfr})). For a coordinate distance $r_0$ between the black hole and the monopole that is large compared to the scale $\Delta r\simeq \eta^{-1}$ of the monopole core\footnote{$\eta$ is the scale of symmetry breaking which we set equal to 1} ($\Delta r << r_0$) the monopole energy may be approximated up to a constant diverging term proportional to $R_{max}$ (independent of $r_0$) as
\be
E\simeq 2\pi \left[r_0^2\;f(r_0)(\frac{\Delta\Phi}{\Delta r})^2+(\frac{\Delta\Phi}{\Delta \theta})^2 + r_0^2 V(0)\right] \Delta r \int_0^{\Delta \theta}\theta d\theta
\label{energymonapprox}
\ee
where $V(0)=\frac{\eta^4}{2}\rightarrow\frac{1}{2}$ and we have set $\sin\theta\simeq\theta <<1$ (see Fig. \ref{fig8}). Also $\Delta \theta \simeq \frac{\Delta r}{r_0}$ is the angular scale of the global monopole core as seen from the center of the black hole. Setting $\Delta \Phi \simeq \eta \rightarrow 1$, $\Delta r \simeq \eta^{-1} \rightarrow 1$ and
\be
\int_0^{\Delta \theta}\theta \delta \theta =\frac{\Delta \theta^2}{2}\simeq \frac{\Delta r^2}{2r_0^2}\rightarrow \frac{1}{2r_0^2}
\label{inttheta}
\ee
we find an approximate expression for the gravitational interaction energy between monopole and black hole as
\be 
E\simeq \pi \left[\frac{5}{2}-\frac{2m}{r_0}-\frac{\Lambda}{3} r_0^2 \right]
\label{energymonapprox2}
\ee
leading to a gravitational force on the monopole from the black hole of the form
\be 
F=-\frac{\partial E}{\partial r_0}=-\frac{2\pi G m \eta}{r_0^2}+\frac{2\pi \Lambda \eta}{3} r_0
\label{forceonmon}
\ee
where we have restored $\eta$ and $G$ for clarity. Thus the monopole behaves like a point test particle with mass equal to $2\pi \eta$. As in the case of a test particle, the force consists of an attractive component due to the black hole mass (as expected from the equivalence principle) and a repulsive component from the cosmological constant. There is an unstable equilibrium point at $r_0=(\frac{3Gm}{\Lambda})^{1/3}$.

\begin{figure}[!t]
\centering
\vspace{0cm}\rotatebox{0}{\vspace{0cm}\hspace{0cm}\resizebox{0.49\textwidth}{!}{\includegraphics{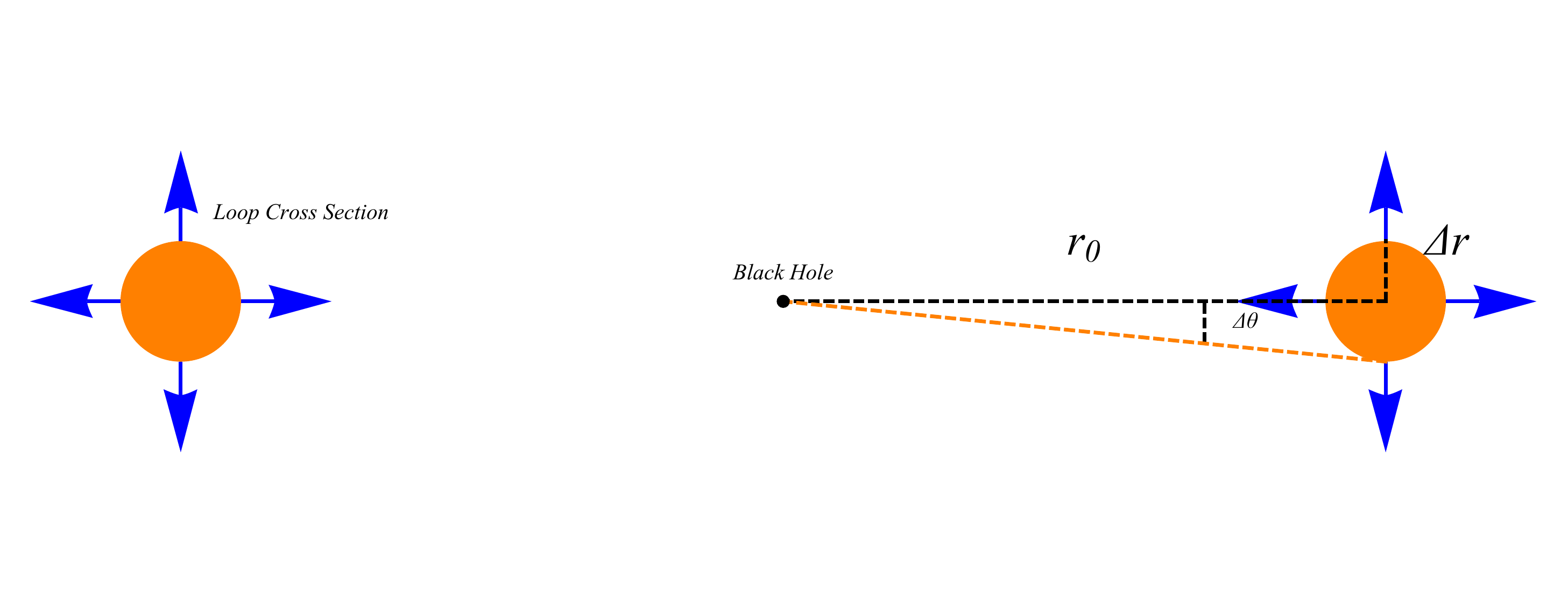}}}
\caption{Cross section view of the global string loop with the central black hole.}
\label{fig9}
\end{figure}
\subsection{Global String Loop in a S-dS spacetime}
\label{sec:Section 4a}

The global field configuration  corresponding to a global string loop is shown in Fig. \ref{fig9}. In this case the energy density is concentrated at $\theta\simeq \frac{\pi}{2}$ rather than $\theta\simeq 0$ and the axial symmetry remains. Setting $\sin\theta \simeq 1$ in eq. (\ref{energymon1}) and using similar approximations and arguments as in the case of the global monopole we find an approximate expression for the energy of the loop which up to a constant diverging term (independent of $r_0$) is of the form
\begin{widetext}
\be
E\simeq 2\pi \left[r_0^2 \; f(r_0)(\frac{\Delta\Phi}{\Delta r})^2+(\frac{\Delta\Phi}{\Delta \theta})^2 + r_0^2 V(0)\right] \Delta r \Delta \theta \simeq 2\pi\left(\frac{5}{2}r_0-2m-\frac{\Lambda}{3}r_0^3\right)
\label{energyloopapprox}
\ee
\end{widetext}
The corresponding effective force is obtained as
\be 
F\equiv -\frac{\partial E}{\partial r_0}\simeq -5\pi G\eta^2 +2\pi G\eta^2 \Lambda r_0^2
\label{forceloop}
\ee
where we have restored $G$ and $\eta$ for clarity.
The first term is an attractive term due to the loop tension while the second term is the repulsive term due to the cosmological constant. The black hole mass does not contribute to the effective force in this case. As in the case of the spherical domain wall we anticipate the existence of an unstable static solution for $r_0\simeq \sqrt{\frac{5}{2\Lambda}}$.

\section{Conclusion}
\label{sec:Section 5}
The main results of this analysis can be summarised as follows:
\begin{itemize}
\item
Derrick's theorem can be violated in the presence of a non-trivial gravitational background. In fact rescaling arguments indicate that static scalar field configurations can exist in the presence of a S-dS background metric.
\item
A spherical domain wall in a S-dS background metric is subject to three types of potentials: a potential that favours contraction due to its tension, a potential that favours expansion due to the central black hole mass and a potential that favours expansion due to the cosmological constant. Expansion occurs for domain wall radius $r_0$ larger than a critical radius $r_{0crit}$ while contraction occurs  $r_0<r_{0crit}$. This result has been demonstrated both analytically and numerically.
\item
A global monopole in a S-dS background metric is subject to two types of potentials: an attractive potential due to the central black hole mass and a repulsive potential due to the cosmological constant. Repulsion dominates for a monopole coordinate distance $r_0$ larger than a critical distance $r_{0crit}$ while attraction occurs  $r_0<r_{0crit}$.
\item
A global string loop in a S-dS background metric with a central black hole is subject to two types of potentials: an attractive potential towards the central black hole due to its tension and a repulsive potential due to the cosmological constant. Repulsion dominates for a loop radius $r_0$ larger than a critical distance $r_{0crit}$ while attraction occurs  $r_0<r_{0crit}$.
\end{itemize}
In all three cases of global defects interacting with a S-dS black hole we anticipate the existence of unstable static solutions for a distance (radius) from the black hole where the above effective forces vanish. 

Interestingly, the mass of the central black hole acts with a repulsive force towards the spherical domain wall but with an attractive force towards a global monopole. This difference does not necessarily lead to violation of the equivalence principle since global defects are nonlocal objects. The cosmological constant acts in a repulsive manner in all three cases of defects.

Interesting extensions of the present analysis include the following:
\begin{itemize}
\item
Extensive numerical simulations of the evolution of a global defects in the vicinity of a S-dS black hole including the cases of global string loops and global monopoles. Such simulations can lead to a  more detailed study of the interaction potentials and the investigation of scattering of global defects from a S-dS black hole.
\item
Consideration of more general background metrics including for example a black hole-monopole or a Kerr background.
\item
Derivation of the predicted gravitational wave features induced by the motion of global defects in nontrivial gravitational backgrounds. Particularly interesting could be the possible gravitational wave and $\gamma$-ray bursts created during the merging of global monopoles with black holes.
\item
Consideration of alternative coordinate systems that may allow the more detailed investigation of the scalar field instabilities that appear to develop beyond the black hole and the cosmological horizons. This will also allow the study of the fate of the global defects as they cross the black hole horizon.
\item
Investigation of the deformation induced on global defect configurations by black holes located at a distance $r_0$ from the center of the defects.
\end{itemize}

{\bf Numerical Analysis Files:} The Mathematica file used for the numerical analysis of this study and for the construction of the Figures may be downloaded from \href{http://leandros.physics.uoi.gr/defects-gravity}{this url}.


\raggedleft
\bibliography{bibliography}

\end{document}